\documentclass[pre,showpacs,preprint,superscriptaddress]{revtex4-1}
\usepackage{amssymb,graphicx,amsbsy,amsmath}

\begin{document}
\title{Internal energy density of the critical three-state Potts model on
the kagome lattice}
\author{Seung Ki Baek}
\email[Corresponding author, E-mail: ]{seungki@kias.re.kr}
\affiliation{School of Physics, Korea Institute for Advanced Study, Seoul
130-722, Korea}
\author{Harri M\"akel\"a}
\affiliation{QCD Labs, COMP Centre of Excellence, Department of Applied
Physics, Aalto University, P.O. Box 13500, FI-00076 AALTO, Finland}

\begin{abstract}
It has been conjectured that the internal energy density of the
Potts model on a semi-infinite strip with a width $L$ does not have any
finite-size corrections at the critical point $K=K_c$.
By factorizing the transfer matrix for the kagome lattice with larger
widths, we have found that this conjecture is not
correct in that the internal energy density slightly varies with $L$ at the
critical point. From this size dependence of the internal energy density,
we obtain an upper bound as $K_c < 1.0565615$, which
is close to a recent estimate $K_c^{\rm JS} = 1.0565600(7)$ by
Jacobsen and Scullard [arXiv:1204.0622]. We also obtain a lower bound as $K_c
> 1.0560$ by calculating the correlation length along the strips.
\end{abstract}

%64.60.De   Statistical mechanics of model systems
%05.70.Jk   Critical point phenomena
%11.25.Hf 	Conformal field theory, algebraic structures

\maketitle

\section{Introduction}

The $q$-state Potts model
has served as a paradigmatic model in classical
statistical physics since it allows to obtain exact and nontrivial analytic
results.
The model is defined by the following Hamiltonian,
\begin{equation}
H = -J\sum_{\left< ij \right>} \delta(S_i, S_j),
\end{equation}
where $J$ is an interaction constant,
the summation runs over the nearest neighbors,
$\delta$ is the Kronecker delta function, and
spin $S_i$ at site $i$ can take a value $0, \cdots, q-1$.
For $q \le 4$, the Potts model on a
two-dimensional (2D) lattice undergoes a continuous phase transition at a
certain inverse temperature $\beta = \beta_c (q)$. It is not a universal
value but depends on the underlying lattice structure. One might say that
locating $\beta_c$ is therefore of somewhat less importance than
understanding universal behavior independent of lattice structures,
but it can be understood as a question
of decisive factors characterizing the critical point when the lattice
symmetry is not high enough, which is still under debate.
In a classical review article~\cite{wu}, the critical point of the
Potts model on the kagome lattice is actually listed as the second unsolved
problem about the Potts model.
Wu's conjecture~\cite{wu} suggests that by solving the following polynomial,
\begin{equation}
v^6 + 6v^5 + 9v^4 - 2qv^3 - 12qv^2 - 6q^2v - q^3 = 0,
\label{eq:wu}
\end{equation}
where $v \equiv e^{K} - 1$ with coupling strength $K \equiv \beta J$,
one can get the critical
point $K_c \equiv \beta_c J$. For $q=3$, the prediction is $K_c \approx
1.056494$, for example.
Equation~(\ref{eq:wu}) is undoubtedly a good
prediction,
but slightly differs from Monte Carlo calculations~\cite{ziff} and series
expansion results~\cite{jensen}. In a recent article~\cite{potts}, this
problem was tackled by using an idea that certain spin models on an infinite
strip with a finite width $L$ do not show finite-size corrections in the
internal energy density $u_L$~\cite{wosiek}. By solving transfer matrices
for two smallest widths $L=1$ and $2$ [see Fig.~\ref{fig:kagome}(a) and
\ref{fig:kagome}(b)] and comparing the resulting internal energy densities,
we conjectured
$K_c^{\rm conj}(q=3)=1.0565094269290\ldots$ and
$K_c^{\rm conj}(q=4)=1.1493605872292\ldots$ using
finite-size scaling arguments.
It was also suggested that one could directly check these conjectured values
by calculating the internal energy density for a strip with $L=3$ shown in
Fig.~\ref{fig:kagome}(c): If the calculated value for $q=3$ deviates from
$u_{L=3}^{\rm conj}(q=3)=-1.6295437063996$ given in Ref.~\cite{potts},
our determination based on the assumption of $L$-independent
$u_L(K_c)$ must be incorrect. In this work, we carry out this calculation by
using the factorization technique~\cite{blote1,*blote2,*foster}
and report that $u_L$ at $K_c^{\rm conj}(q=3)$ is different from
the conjectured value $u_L^{\rm conj}(q=3)$ given above. It means that the
absence of finite-size corrections assumed in Ref.~\cite{potts} is not true
in general except for $q=2$. This was recently
pointed out by Jacobsen and Scullard~\cite{scullard}:
They mentioned the existence of finite-size corrections for $q \neq 2$,
and our conjectured values are not compatible with their results
obtained by improving the polynomial in Eq.~(\ref{eq:wu}) systematically
with larger subgraphs of the kagome lattice.
So the conjecture in Ref.~\cite{potts} is disproved, which is the main
result of this work. The second is that it is nevertheless possible to find
bounds for $K_c$ by using the size dependence of numerical results. Our
upper bound in this work is $K_c < 1.0565615$, and a lower bound is obtained
as $K_c > 1.0560$, which are consistent with the value $K_c^{\rm JS} =
1.0565600(7)$ estimated by Jacobsen and Scullard~\cite{scullard}.

\section{Method and Result}

\begin{figure}
\begin{center}
\includegraphics[width=0.15\textwidth]{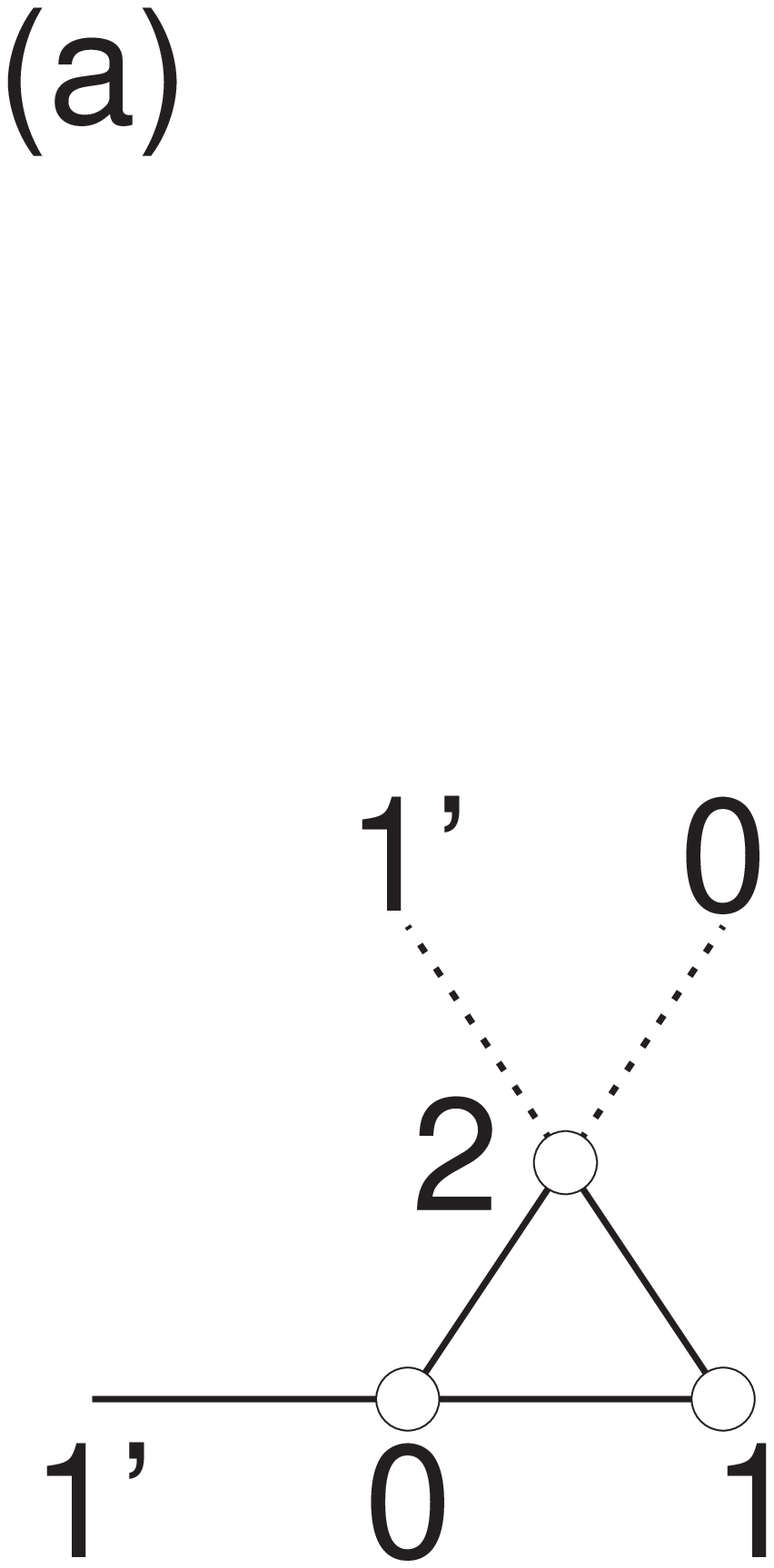}
\includegraphics[width=0.15\textwidth]{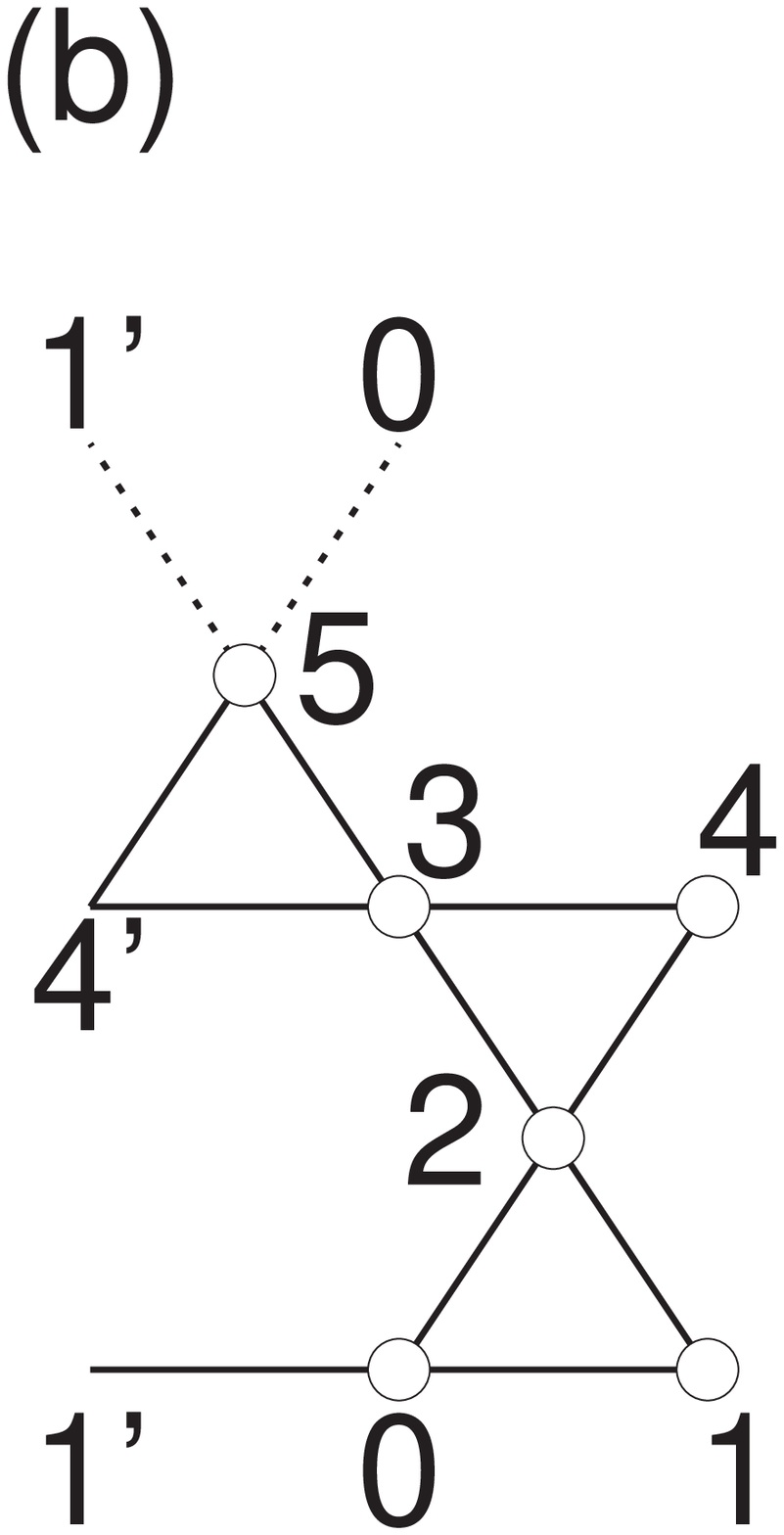}
\includegraphics[width=0.15\textwidth]{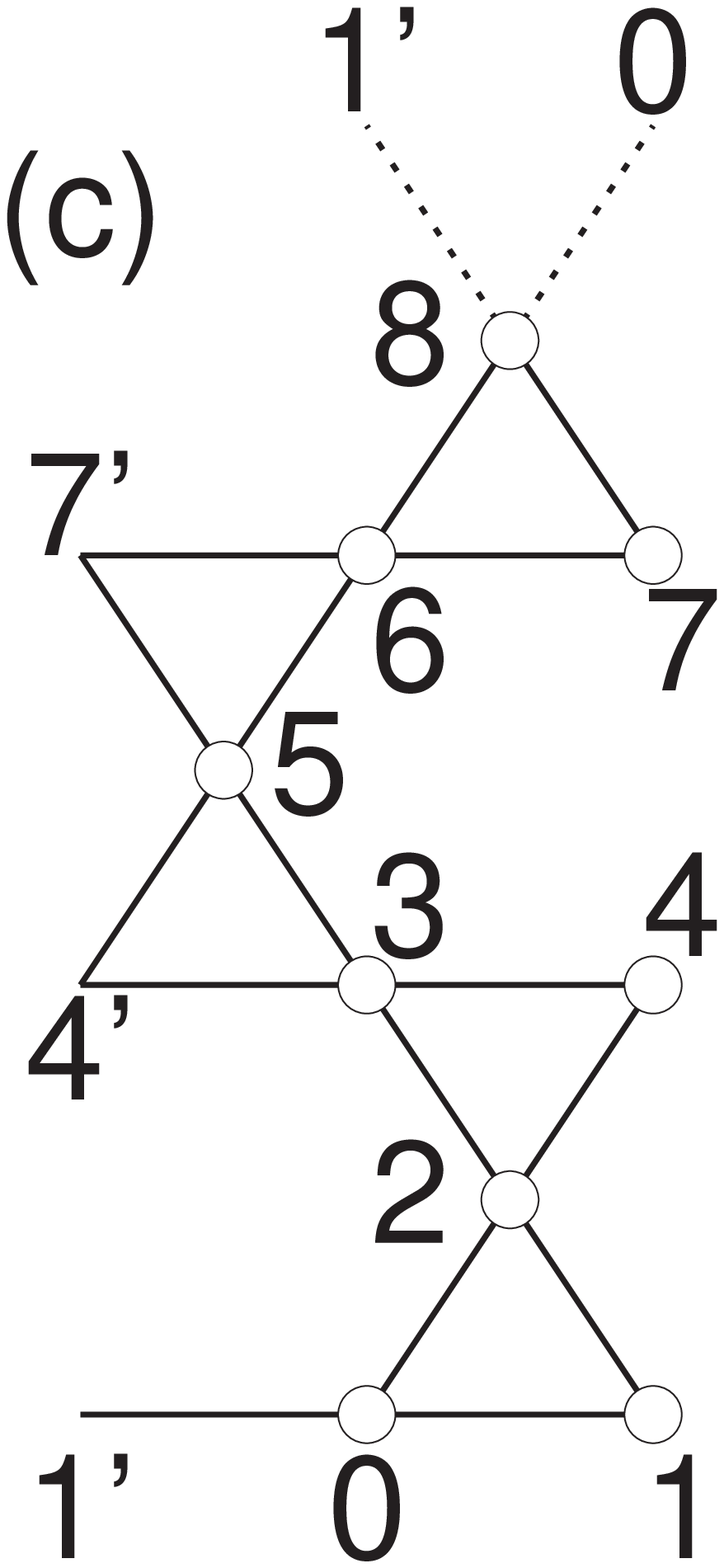}
\end{center}
\caption{ Unit cells of the kagome lattice with different widths (a) $L=1$,
(b) $L=2$, and (c) $L=3$. The dotted lines mean the periodic boundary
condition so that $1'$ and $0$ on the top are connected to $1'$ and $0$ at the
bottom, respectively. }
\label{fig:kagome}
\end{figure}

\begin{table}
\caption{One possible correct order of adding spins for $L=3$
[Fig.~\ref{fig:kagome}(c)]. A link between spins $i$ and $j$ is
denoted by $(i,j)$.}
\begin{tabular*}{0.45\textwidth}{@{\extracolsep{\fill} } cc}
\hline
spins added & links added \\\hline
0 & (0,1')\\
2 & (2,0)\\
3 & (3,4'), (3,2)\\
5 & (5,4'), (5,7'), (5,3)\\
4 & (4,3), (4,2)\\
6 & (6,7'), (6,5)\\
7 & (7,6)\\
8 & (8,1'), (8,0), (8,6), (8,7)\\
1 & (1,0), (1,2)\\\hline
\end{tabular*}
\label{table:order}
\end{table}

It is straightforward in principle to construct a transfer matrix $T$
describing the three-state Potts model on an infinite strip with a width $L$.
The problem is that it gets too large as $L$ grows. Therefore, we need to
factorize $T$ into sparse matrices~\cite{blote1,*blote2,*foster} and then use
the power iteration method to get the largest eigenvalue and the
corresponding left and right eigenvectors.
The factorization means that a transfer matrix of a unit cell of width $L$
as in Fig.~\ref{fig:kagome} is equivalent to a product of $3L$ sparse matrices
each of which describes adding a single spin. The advantage of the
factorization lies in the fact that multiplying a vector by the sparse
matrices one after another is much faster than dealing with the original
dense matrix at once, because only a few elements of the vector are affected
by a sparse matrix.
Note that we have to be careful about the
order of adding spins under the periodic boundary
condition along the vertical direction, because each spin addition
amounts to replacing a spin value in a layer by a new one. For example, in
Fig.~\ref{fig:kagome}(a), adding spin $1$ replaces the old value of spin
$1'$, eliminating its information.
For that reason, one possible correct order for $L=1$ is
$0 \rightarrow 2 \rightarrow 1$, where spin $1$ is added after spin $2$,
because spin $2$ should have a link to spin $1'$
before spin $1$ replaces it, according to the periodic boundary
condition [see the dotted line between spin $2$ and spin $1'$ in
Fig.~\ref{fig:kagome}(a)].
For the same reason, one correct order for $L=2$ is $0 \rightarrow 2
\rightarrow 3 \rightarrow 5 \rightarrow 4 \rightarrow 1$, and
Table~\ref{table:order} shows a possible order for $L=3$ in detail. 
This approach can be extended to larger $L$'s in a straightforward manner.
For direct comparison, we have chosen the same transfer direction as in
Ref.~\cite{potts}.
It is notable that Ref.~\cite{scullard} also carried out transfer matrix
calculations to assess their approximation but in a different transfer
direction.

As we multiply an arbitrary vector by the transfer matrix $T$ over and over
again, it almost surely converges to an
eigenvector with the largest eigenvalue $\lambda_1$ of $T$. Once we have
obtained the left eigenvector $\mathbf{v}_L$ and the right eigenvector
$\mathbf{v}_R$ corresponding to $\lambda_1$, we are able to get
$u_L = -(3L\lambda_1)^{-1} \partial \lambda_1 / \partial \beta =
-(3L\lambda_1)^{-1} \mathbf{v}^{\ast}_L \cdot(\partial T / \partial \beta)
\cdot\mathbf{v}_R$, where $\ast$ means complex conjugate transpose. The
factor of $3L$ appears since it is the number of spins for constructing $L$
layers as depicted in Fig.~\ref{fig:kagome}. The numerical error $\delta u_L$ in
the internal energy density increases with $L$ and it is estimated as
$\delta u_L \sim O(10^{-11})$ for $L=5$ and $O(10^{-9})$ for $L=6$.
For $L=3$,
the internal energy density at $K_c^{\rm conj}(q=3)$ obtained by this
method reads $u_{L=3}^{\rm power} = -1.62949\ldots$, which clearly deviates
from $u_{L=3}^{\rm conj}=-1.62954\ldots$ conjectured in Ref.~\cite{potts},
disproving the conjecture on the lack of finite-size corrections in $u_L
(K_c)$.

\begin{figure}
\begin{center}
\includegraphics[width=0.45\textwidth]{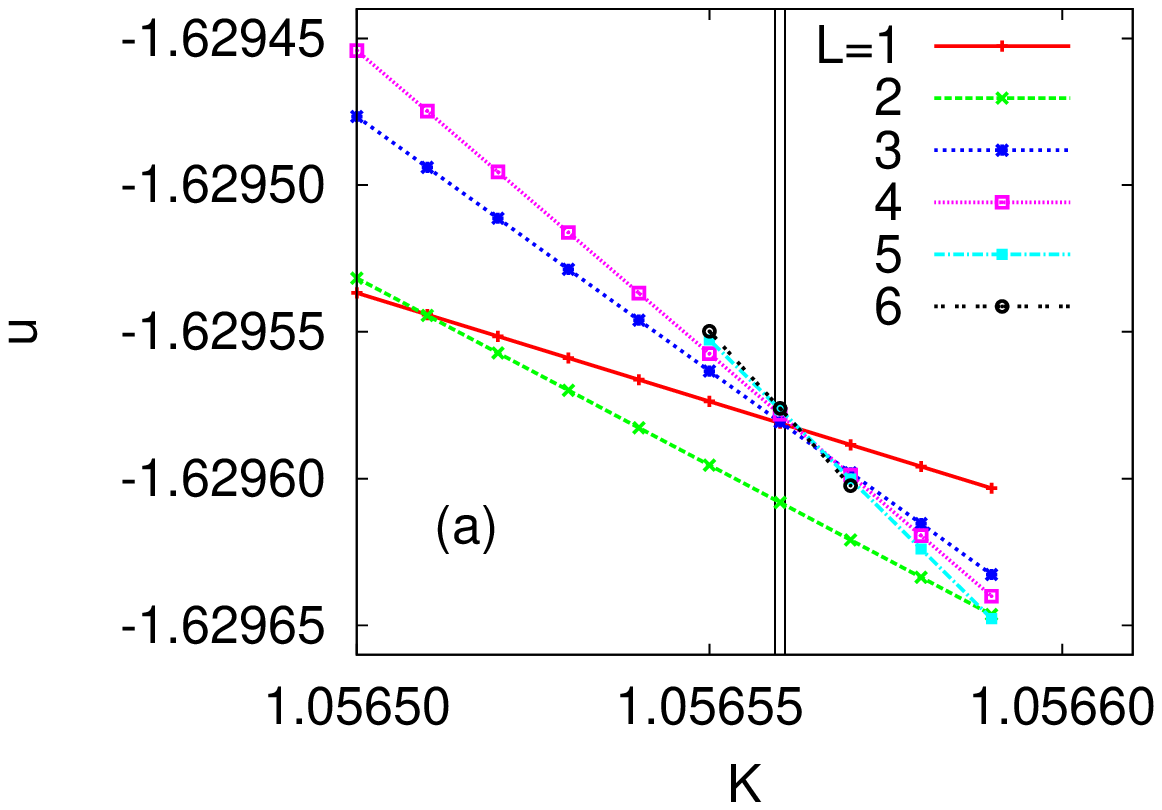}
\includegraphics[width=0.45\textwidth]{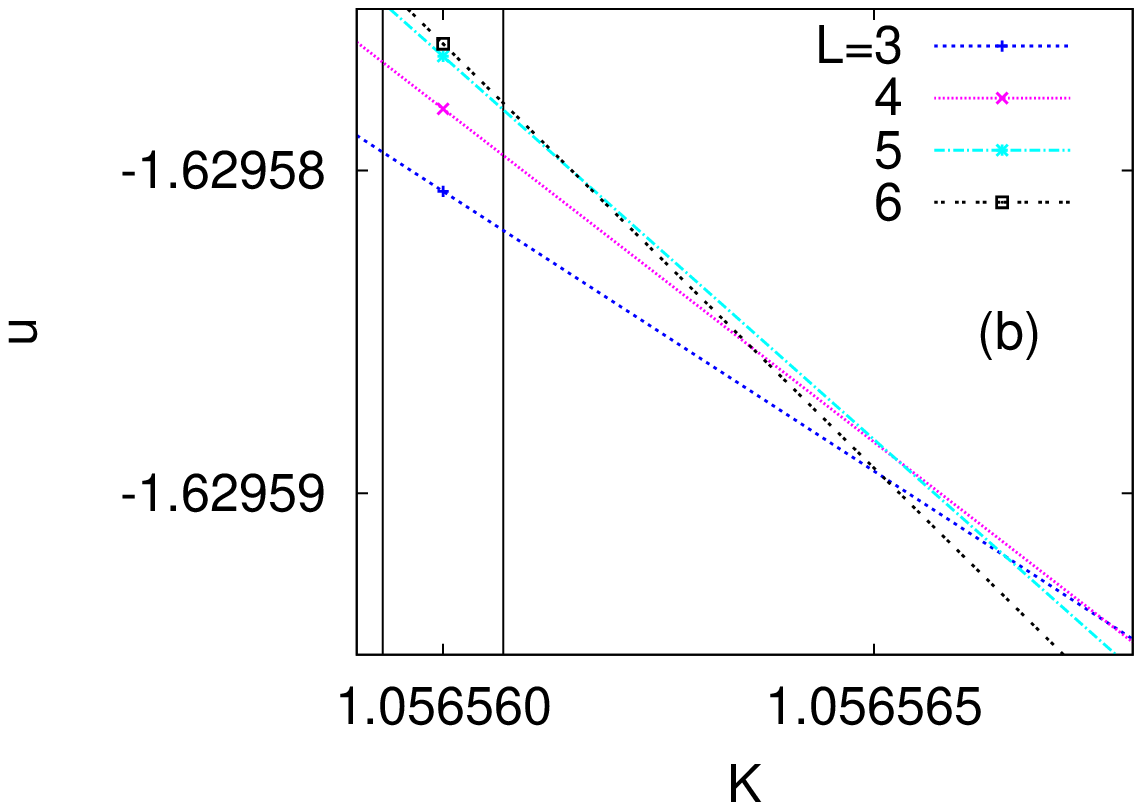}
\includegraphics[width=0.45\textwidth]{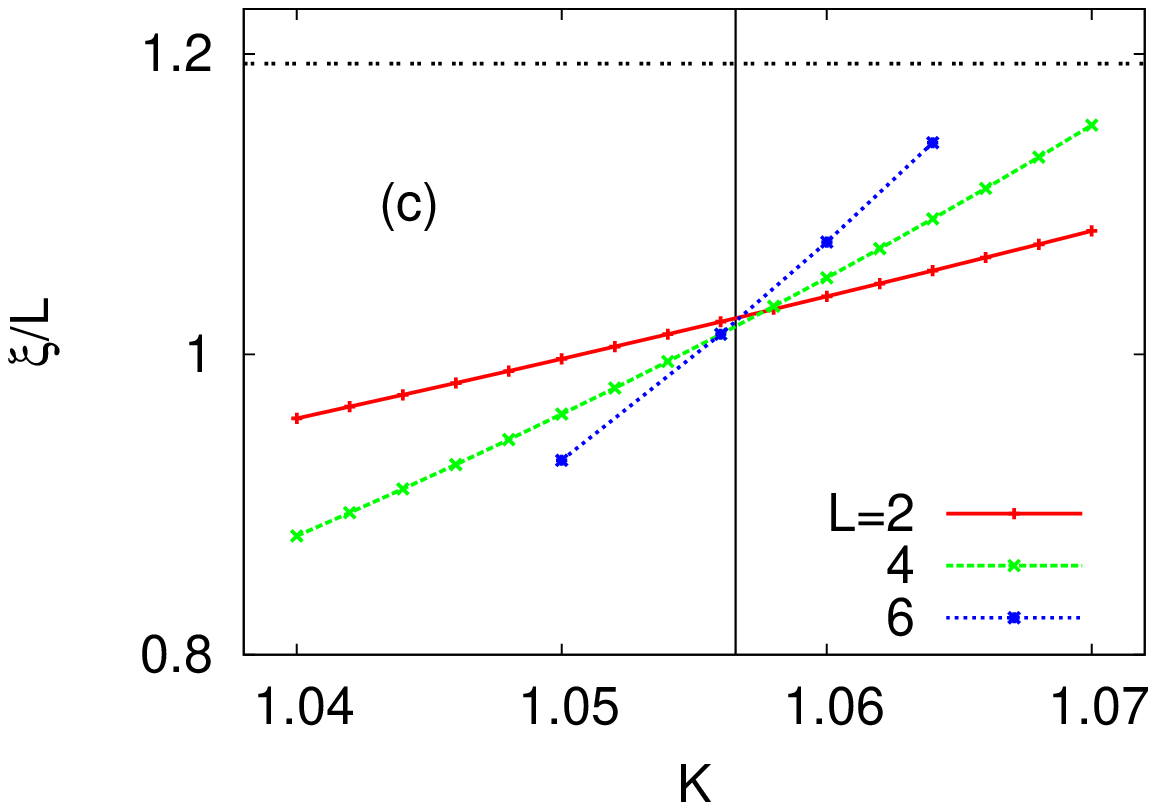}
\includegraphics[width=0.45\textwidth]{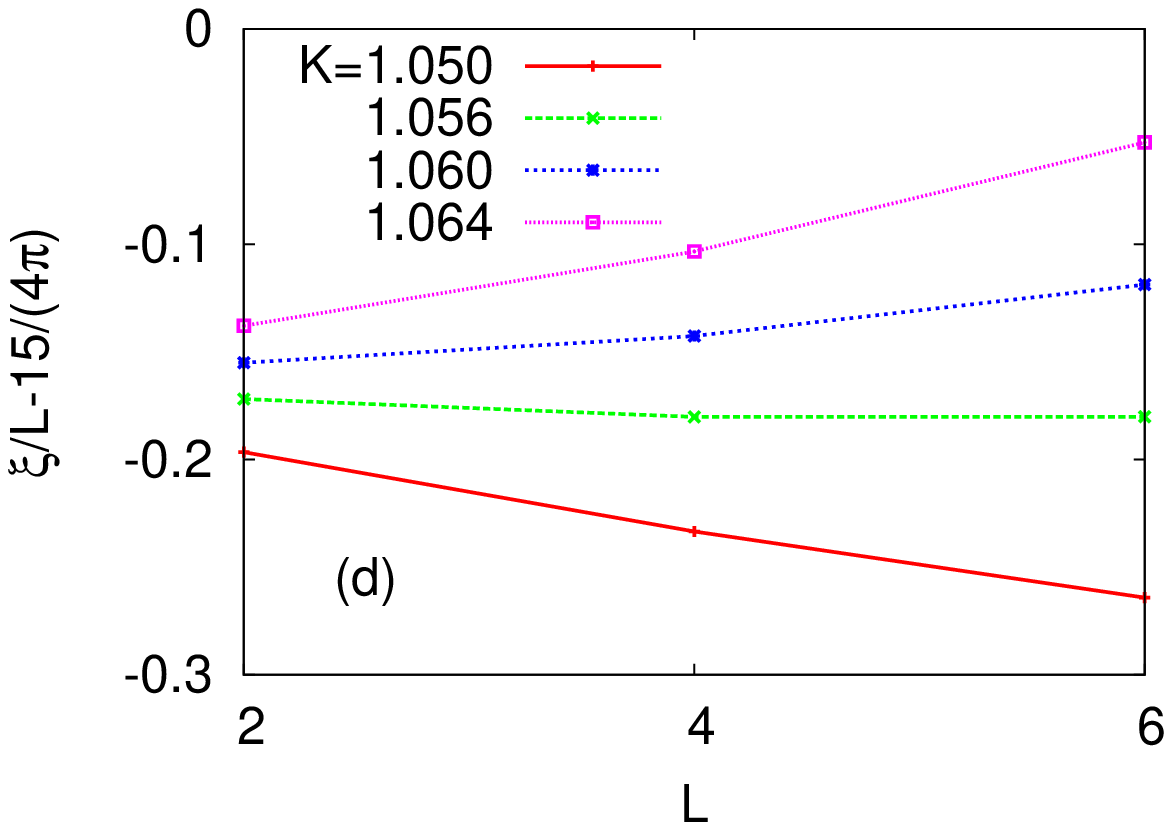}
\end{center}
\caption{(Color online) (a) Internal energy density as a function of $K$ for
each strip width $L$ and (b) its zoomed view.
(c) Correlation length $\xi$ divided by $L$, where the dotted horizontal line
means $1/(\pi\eta) = 15/(4\pi)$.
The vertical lines in panels (a) to (c) indicate the range of $K_c$
estimated by Jacobsen and Scullard~\cite{scullard}.
(d) As $L$ increases, the convergence of
$\xi/L$ to $15/(4\pi)$ is observed for $K>1.056$.
}
\label{fig:energy}
\end{figure}

Our results up to $L=6$ are depicted in Fig.~\ref{fig:energy}(a).
The line of $L=1$ behaves differently from the
others since the size is too small and should therefore be discarded from
consideration.
The crossing points of lines for $L \ge 2$ 
gradually converge to $K_c$ from above [Fig.~\ref{fig:energy}(b)].
It is therefore plausible that $K_c$ is
located to the left of the crossing between $L=5$ and $L=6$ at $K
\approx 1.0565615$.
Although the size-dependence of $u_L$ is known for large $L$ at the critical
point, our fitting result to estimate $K_c$ from this critical scaling
behavior is inconclusive because sub-leading corrections are not negligible
for such small $L$'s. 
%\begin{equation}
%u_L \approx u_{\infty} + A L^{-(1-\alpha)/\nu},
%\label{eq:size}
%\end{equation}
%where $u_{\infty}$ is the bulk value, $A$ is a constant, and $\alpha=1/3$ and
%$\nu=5/6$ are the scaling exponents of the model.

In order to estimate a lower bound, we use correlation length
$\xi = [\ln(\lambda_1/\lambda_2)]^{-1}$, where $\lambda_2$ is the second
largest eigenvalue of $T$. The second largest eigenvalue can be found by
applying the power iteration method to a vector
orthogonal to $\mathbf{v}_R$, the eigenvector associated with $\lambda_1$.
The numerical error in $\xi$ is estimated as $O(10^{-11})$ for $L=5$ and
$O(10^{-7})$ for $L=6$.
As $L$ increases, it is
known that $\xi/L$ converges to $1/(\pi \eta) = 15/(4\pi) \approx
1.194$ at criticality, where $\eta$ is the critical exponent for the two-spin
correlation function~\cite{cardy}. Since $\xi$ is an increasing function of
$K$ and the lines cross each other below $1/(\pi \eta)$ [see
Fig.~\ref{fig:energy}(c)], the crossing point will approach $K_c$ from
below if $L$ is large enough.
We thereby infer a lower bound as $K_c > 1.0560$, where the
lines of $L=4$ and $L=6$ cross each other. The convergence of $\xi/L$ to
$15/(4\pi)$ also suggests that the borderline is around $K=1.056$
[Fig.~\ref{fig:energy}(d)].
Note that we compare results from
even $L$'s here because the behavior of $\lambda_2$ seems more
sensitive to the boundary condition in the vertical direction than that of
$\lambda_1$.

\section{Summary}
In summary, we have numerically calculated internal energy densities near
$K=K_c^{\rm conj}$ for the three-state ($q=3$) Potts model on the
kagome-type strips with various widths.
The result disproves our previous
conjecture that the internal energy density has no finite-size corrections
at the critical point.
The size dependence suggests $1.0560 < K_c < 1.0565615$,
which is consistent with the recent estimate $K_c^{\rm JS} =
1.0565600(7)$~\cite{scullard}.
 
\acknowledgments
H.M. was supported by the Alfred \mbox{Kordelin} Foundation and the Academy
of Finland through its Centres of Excellence Program (Project No. 251748).
We thank KIAS Center for Advanced Computation for providing computing
resources.

%\bibliographystyle{apsrev4-1}
%\bibliography{kag}
\end{document}